# Modification in structural, dielectric and magnetic properties of La and Nd co-substituted epitaxial BiFeO$_3$ thin films

Anju Ahlawat[1]*, S. Satapathy[1], V. G. Sathe[2], R. J. Choudhary[2], M. K. Singh[1], Ravi Kumar[3], T. K. Sharma[3], P.K.Gupta[1]

[1] Nanocrystalline Functional Materials Group, Laser Materials Development & Devices Division, Raja Ramanna Centre for Advanced Technology, Indore 452013, India

[2] UGC-DAE Consortium for Scientific Research, University Campus, Khandwa Road, Indore – 452 017, India

[3] Semiconductor Laser Section, Solid State Laser Division, Raja Ramanna Centre for Advanced Technology, Indore 452013, India

Corresponding author
*E- mail: anjahlawat@gmail.com
Phone: 91 731 2488660/8658
Fax: 91 731 2488650

**Abstract**

The influence of La and Nd co-substitution on the structural and magnetic properties of $BiFeO_3$ (BFO) thin films was examined. Epitaxial thin films of pure and, La and Nd co-doped BFO on the $SrRuO_3$ buffered single crystal $SrTiO_3$ (001) substrate were deposited using pulsed laser deposition. The structural change in co doped La and Nd BFO thin films which was caused by the changes of force constant in the crystal lattice induced by ionic radii mismatch was investigated. Raman spectroscopy studies manifest the structural change in doped BFO films from rhombohedral to monoclinic distorted phase which is induced by the co substitution of La and Nd. Room temperature magnetic hysteresis curves indicated that saturation magnetization is enhanced in the doped film with saturation magnetization of ~20 emu/cm$^3$. The dielectric and magnetic properties are effectively improved in BLNFO films compared to pure BFO thin films.

**1. Introduction**

Magnetic ferroelectrics, having magnetic and ferroelectric ordering simultaneously have attracted much attention from both fundamental and practical point of view [1]. Recently, ferroelectromagnetism is the subject of intensive investigation because of the potential to offer a wide range of application, including the emerging field of spintronics, memory and data storage media [2]. The BiFeO$_3$ (BFO) is to be known the only material which can be used at room temperature because it has high electric ordering temperature ($T_c$ ~ 1103 K) and a high magnetic ordering temperature ($T_N$ ~ 643 K) [3]. BFO possesses a rhombohedrally distorted perovskite structure with space group R3c which allows antiphase octahedral tilting and ionic displacements from the centrosymmetric positions about and along the same (111) cubic direction. It exhibits weak magnetism at room temperature (RT) due to the residual moment from a canted spin structure [4-6]. In spite of having a very high ferroelectric and Neel transition temperature, there are several problems such as large leakage current and spatially non uniform magnetic structure which leads to the cancellation of macroscopic magnetization, have restricts the application of BFO in devices at room temperature. According to Illiev *et al* [7] the rhombohedral R3*c* structure with polarization along [111] direction gives many challenges in controlling meaningful multifunctionality of this material. Therefore efforts are on to realize other structures with tetragonal symmetry. It is predicted that tetragonal symmetry will resolve many of the problems and will have potential for applications.

It is possible to modify the rhombohedral symmetry by chemical substitution and/or by elastic constraints in thin films. It is reported that the chemical substitution modulate the spin structure of BFO [3] or by epitaxial constraints its multiferroics properties can be improved [7-9]. The multiferroic properties and the structure of La-doped BFO thin films deposited by pulse laser deposition (PLD) and chemical solution deposition (CSD) have been studied extensively [10,11]. Dongeun Lee *et al* [10] reported that La doping modified the structure of BFO films from the monoclinic state to the tetragonal state. Das *et al* [11] showed that the leakage current decreased after La doping in the BFO films grown on Pt/TiO2/SiO2/Si substrates by PLD. Uchida *et al* [12] reported that the crystal anisotropy and the Curie temperature of BFO degraded continuously with increasing contents of La doping. Recently, increasing attention has been paid to the co-doped films to further enhance the multiferroic properties [13, 14].

In this article, we report the structural changes observed in BFO thin films induced by La and Nd co-doping on the A-site which also has a subsequent impact on the structure, dielectric and magnetic properties of these films. The idea of simultaneous La and Nd substitution is based on the interesting observation; recently, Huang *et al* [15] reported that the ferroelectric polarization and saturation magnetization of the BFO films were enhanced by appropriate Nd doping (only upto 10 %) whereas in $Bi_{1-x}La_xFeO_3$, the increasing concentration of La, leads to the structural phase transition at $x < 0.2$, which suppress spatially inhomogeneous spin-modulated incommensurate structure of BFO and resulted in the improved magnetization [16].

The most important issue of local symmetry changes due to doping and elastic strain in films can be very effectively probed by Raman scattering [17]. In this framework, pure BFO as well as co-doped $Ba_{0.8}La_{0.1}Nd_{0.1}FeO_3$ (BLNFO) film are prepared and extensively studied by polarized Raman spectroscopy with an expectation that the structural, ferroelectric and ferromagnetic properties of BFO films can be improved by La and Nd co-substitution.

## 2. Experimental

The BFO and BLNFO thin films were grown on the single crystal $SrTiO_3$ (STO) (001) substrate by pulsed laser deposition using KrF Excimer (•=248 nm). The parent compounds i.e. BFO, $SrRuO_3$ and BLNFO (used as ablation targets for films) were synthesized using conventional solid state reaction methods which are highly pure in nature as described in our other reports [18]. The SrRuO$_3$ (SRO) was used as the bottom electrode to facilitate the electric measurements. In order to deposit SRO layer, the chamber was first evacuated up to a base pressure of $5\times10^{-7}$ mbar and the depositions were carried out in the presence of oxygen partial pressure of 100 mtorr. An optimized substrate temperature of 600°C, laser repetition rate of 3 Hz and a laser density of 2 J/cm$^2$ were maintained during deposition of the 50 nm thick SRO layer. Subsequently, the substrate (with SRO layer) was heated to 650°C and the oxygen partial pressure was maintained at 50 mTorr while keeping other deposition conditions same, BFO films of ~200 nm thickness were deposited. After deposition, the samples were cooled to room temperature in 1 atm O$_2$ ambient.

The crystal structure of the films was examined by normal X-ray diffraction (XRD) with a Cu K$\alpha$ radiation source. The degree of in-plane orientation was assessed by examining phi scans and reciprocal space maps which were recorded by high Resolution X-Ray Diffraction (HRXRD) technique. HRXRD experiments were carried out using Panalytical X'Pert PRO MRD system where a hybrid monochromator was used to provide CuK$\alpha_1$ monochromatic X-ray beam ($\lambda$=1.5405Å) with a beam divergence of ~ 20 arcsec. HRXRD phi scans were done with a 0.75° open detector parallel to the scattering plane [19]. In order to record the reciprocal lattice maps a Ge (2 2 0) three-bounce monochromator was placed in front of the detector to ensure an acceptance angle of ~ 12 arcsec [20, 21]. The surface morphology of the films was obtained by the field emission scanning microscope (FESEM) and Atomic Force microscope (AFM). The thicknesses of the films were measured using FILMETRICS (F-20) thickness measurement system. Raman measurements were performed in backscattering geometry using a Ar$^+$ excitation source having wavelength 488 nm coupled with a Labram-HR800 micro Raman spectrometer equipped with a ×50 objective, appropriate edge filter and peltier cooled charge coupled device detector. The magnetic properties of the films were investigated with Superconducting QUantum Interference Vibrating sample magnetometer (SQUID-VSM) with the magnetic field applied parallel to the films plane.

## 3. Result and discussion

### 3.1. Structural and morphological studies

Figure 1 shows X-ray diffraction pattern of BFO and BLNFO films deposited on $SrRuO_3$ buffered $SrTiO_3$ (001). For both the films only strong peaks corresponding to (*002*) plane of BFO, SRO are observed, which shows highly oriented growth of $SrRuO_3$ phase along with BFO phases along (*002*) orientation. Compared to the pure BFO film, a shift of (*002*) peak towards higher 2θ angle is observed in BLNFO film. For pure BFO film, SRO peak is well separated from film as well as substrate peak (as shown in the inset of figure 1) however, in case of BLNFO film SRO peak is overlapped with film peak due to the very close lattice parameter of SRO and BLNFO. Figure 2(a) shows phi scan of BFO film for (022) reflection. It shows four peaks which are separated by 90 degree from each other. Reciprocal space map for the same reflection, as shown in Fig. 2(b), confirms that the in-plane lattice parameter of BFO and SRO are almost the same as that of STO. Phi scan as well as reciprocal space map across (022) plane confirms in plane epitaxy of the BFO films. In case of BLNFO film, the phi scan and reciprocal space maps could not be recorded which might be due to the reason that for BLNFO film, 2θ value becomes close to STO and moreover, due to the overlapping of BLNFO and SRO peak (as shown in XRD pattern). The out of plane lattice parameters (when pseudo cubic unit cell is considered) are found to be 4.022 Å and 3.970 Å for the BFO and BLNFO films respectively. The out of plane unit cell parameters of the films were found to be larger than that of the bulk BFO and BLNFO samples [18]. It indicates the presence of tensile strain in out of plane direction for both the films. The lattice parameter for BFO is found to be slightly different than that of BLNFO. One possible explanation may be given in terms of the difference in the ionic radius of dopant, the ionic radius of $Nd^{3+}$, $La^{3+}$, and $Bi^{3+}$ are 0.983, 1.032 and 1.03 Å, respectively. Similar to the other reports, pure BFO (001) film adopt rhombohedral crystal structure [22] but the atomic substitution in BFO films affects the lattice parameter as well as the crystal structure [23-25].

However, it is difficult to examine the change in crystal structure of the films using XRD. By comparison, Raman spectroscopy has been proved to be very excellent techniques for the evolution of structure more explicitly. Figure 3 shows the two dimensional AFM of BFO and BLNFO thin films over $5 \times 5$ $\mu m^2$ area. AFM studies revealed that both the films appear to be dense, crack-free. However BFO and BLNFO films exhibit different textures. Pure BFO film has very smooth surface and made up of homogeneous grains with distinct grain boundaries. Whereas for doped film (BLNFO) homogeneity of the grains get affected and the microstructure of the films appears to be modified. The root mean square roughness of BFO and BLNFO are 1.1 and 3.8 nm respectively. Figure 4 shows field emission scanning electron microscope (FESEM) micrographs for the BFO and BLNFO thin films over 200 nm area. The FESEM results show that both the films were uniformly grown without porosity. The grain size was estimated to be 50-150 nm and ~100 nm for BFO and BLNFO respectively. The inset of fig.4 shows three dimensional SEM images of BLNFO thin films. Both AFM and FESEM studies indicate that the incorporation of La and Nd to BFO significantly affects the surface morphology and the microstructure of the films.

## 3.2. Raman Studies

The room-temperature structure of BFO crystal is a highly distorted rhombohedral perovskite structure with space group *R3c*. There have been several reports on the Raman spectra of BFO obtained on single crystal [26, 27], polycrystalline samples [15, 28] and epitaxial thin films grown on SrTiO$_3$ substrates with different orientation (001, 111) etc [26, 30]. However, the assignment of modes in thin films is a very complicated task because a very small distortion induced during the preparation can lead to different crystal symmetry. Therefore in the present work, polarization dependent Raman measurements were performed to assign the modes and to examine the structure of pure as well as co-doped (La and Nd ions on A site) BFO films. It is worth mentioning that the polarized Raman is able to give local symmetry change that is a very important factor for ferroelectric properties in BFO. All the measurements have been performed in backscattering geometry (incident wave vector parallel to the scattered one) at room temperature. The unpolarized (UP) Raman spectra of STO and BFO/SRO/STO are shown at room temperature in fig. 5. The comparison of spectrum of BFO/SRO/STO with STO spectrum, exclude any Raman contribution from STO (substrate). According to group theoretical calculations, the rhombohedral (*R3c*), tetragonal *(P4mm)* and monoclinic [7] structures of BFO give rise to 13, 10 and 27 distinct Raman-active modes, respectively, *Rhombohedral*, $R3c(C_{3v}) = 4A_1 + 9E$, *Tetragonal P4mm = $4A_1+B_1+5E$, Monoclinic Cc = $13A' + 14A''$*.

Figure 6 shows unpolarized Raman spectra of the pure and doped BFO sample. In the present study BFO thin film revealed intense peaks located at 72, 140, 175, and 220 cm$^{-1}$ corresponds to the $A_1$ modes and the weaker peaks (as marked in fig. 6) at 277, 355, 378, 473, 526, 569 and 616 cm$^{-1}$ which can be assigned to E modes. However, for BLNFO film, numbers of peaks are larger than that of the BFO film. BLNFO film show strong peaks at 71, 138, 170, and 218 cm$^{-1}$ and weaker peaks at 267, 299, 340, 369, 414, 473, 519, 546, 607 and 670 cm$^{-1}$. In order to get a better comparison of BFO and BLNFO films the Raman spectra was recorded at low temperature (80K) and that is shown in fig. 7. The difference of number of Raman modes for BFO and BLNFO films is revealed more clearly at low temperatures. The larger number of peaks (shown by arrow in fig. 7) in the BLNFO film reflects lowering of symmetry when compared to that for BFO films. These weak modes at higher frequencies might be due to the presence of monoclinic distortion as reported in ref. [31]. The spectra of BFO and BLNFO films were fitted with multiple peaks using lorenzian line shape function and shown in fig. 8. For pure BFO film the spectra is well fitted using 13 number of lorengian functions concurrent with rhombohedral structure [7]. The fitting results on the other hand for BLNFO films required larger number of peaks in order to get good fitting. This establishes change in crystal symmetry other than rhombohedral structure for doped film. In order to discern the changes in symmetry due to La and Nd doping in BFO structure, polarization dependent measurements were performed for BFO and BLNFO films. The selection rule for the total number of normal Raman modes (N) in different polarization configuration for rhombohedral (*R3c*), tetragonal (*P4mm*), and monoclinic (*Bb*) crystal structures are given in table 1. The comparison of Raman active mode and assigned symmetries of BFO crystal for the present work and earlier reports is summarized in Table 2.

Figure 9 shows the polarization dependent Raman spectra for BFO and BLNFO films. The polarization dependent measurements were performed with parallel (XX) (the wave vector of scattered light is parallel to the incident light wave vector) as well as cross polarization configuration (XY) (the wave vector of scattered light is perpendicular to the incident light wave vector). Thus the parallel and cross polarization configuration corresponds to $Z(XX)\bar{Z}$ and $Z(XY)\bar{Z}$ geometry respectively (as shown in the inset of the fig. 9) following Porto notations [32]. Figure 9b shows the polarized Raman spectra of STO substrate in which the broad peak at 76 cm$^{-1}$ is retained in all configurations. But in case of BFO thin films (fig. 9a) the peak at 72 cm$^{-1}$ is suppressed in $Z(XY)\bar{Z}$ configuration which indicated that the peak at 72 cm$^{-1}$ is due to BFO thin films. According to the Raman selection rules for rhombohedral structure, only $A_1$ modes are expected to appear in the $Z(XX)\bar{Z}$ configuration and should not be seen in $Z(XY)\bar{Z}$ configuration. On the other hand E modes are allowed in both $Z(XX)\bar{Z}$ and $Z(XY)\bar{Z}$ configuration. Figure 9 shows that the $A_1$ modes suppressed whereas E modes are appeared in $Z(XY)\bar{Z}$ configuration for BFO film which is in accordance with the polarization selection rules for rhombohedral structure [7]. Though there are several reports on BFO films but still there are controversies in literature about the crystal structure of BFO thin films. Singh *et al.* and Palali *et al.* have interpreted the tetragonal P4*mm* and monoclinic B*b* structures, respectively for the BFO/SRO films they have studied [30,31]. However, Illiev et al. [7] insisted that the film thickness 300 and 600 nm are favorable for growth of rhombohedral structure and as reflected in their Raman measurements. The Raman spectra of BFO films were compared to the relaxed BFO-R (80 nm) [30] as well with the Raman spectra reported by others [26, 28]. In present case our polarized Raman results showed selection rules according to rhombohedral R3c structure for strained epitaxial BFO films of 150 nm thickness. Reciprocal space maps (fig. 2(b)) confirmed that the BFO/SRO/STO films are strained.

For BLNFO films, the behaviour in polarization analysis is markedly different when compared to BFO films. Here we see existence of prominent modes in $Z(XX)\bar{Z}$ and $Z(XY)\bar{Z}$ configuration which is not possible in R3c symmetry according to selection rules [7]. Therefore we looked for other symmetries like tetragonal P4mm or monoclinic Bb space group. Our polarization analysis results are similar to the recent report on BFO/SRO/STO (001) films with monoclinic structure [31]. Theoretical calculations for Raman intensity for monoclinic symmetry Bb suggest that for parallel polarization geometry $Z(XX)\bar{Z}$ the intensity of $A^/$ modes should be proportional to $\frac{1}{4}(a+b)^2$ while in cross polarization geometry $Z(XY)\bar{Z}$ the intensity for $A^/$ mode should be proportional to $\frac{1}{4}(a-b)^2$. From structural point of view unless the monoclinic distortion is large the value of a will be very close to b. This should result in nearly zero intensity for the $A^/$ modes in cross polarization geometry. The intensity of A1 modes for rhombohedral symmetry in cross polarization mode is also proportional to $(a-b)^2$ [30] and in this case a is expected to be equal to b giving zero intensity for these modes in cross polarization geometry. Our results on pure BFO films show zero intensity for these modes in cross polarization geometry (see fig. 9a) while a small but noticeable intensity for BLNFO film. The observation of intensity for these mode for BLNFO film again suggest that it does not have rhombohedral symmetry and the monoclinic distortion is enhanced due to doping giving possibly different values for a and b Raman tensor elements. This is consistent with our XRD results where the lattice parameters for the two films are significantly different.

It has been reported that structure has been changed from tetragonal to monoclinic with increasing the thickness of the BFO films from 70 to 240 nm [33]. However, in the present case the thickness for both films is same; hence the possibility of structural transformation due to thickness variation is ruled out. The modification in structure is due to the co doping of La and Nd on the A site in BFO films. Apart from the structural changes observed in doped film, red shift is observed in the Raman bands frequencies for BLNFO films as compared to the pure BFO film. It is known that the increase or decrease of the Raman frequency shift is associated with the tensile/compressive stress existing in the film [34, 35]. The larger the tensile stress that exists in the film, the higher frequency the Raman shift appears. This indicated that BFO film is under tensile strain in comparison to the BLNFO film which is attributed to the difference in the percentage of lattice mismatch between substrate and BFO/BLNFO. However, the magnitude of the Raman frequency shift is not same for all the modes, a large shift is observed in the frequency of $A_1$ mode at 170 cm$^{-1}$ as compared to the other modes (as shown in the fig. 6). It was observed that the line width (FWHM) of $A_1$ mode at 170 cm$^{-1}$ mode for BLNFO film is smaller than that of BFO film.

A first-principles calculation predicted that Bi atoms responsible for the ferroelectric phase transition participate in low-frequency modes up to 167 cm$^{-1}$ while oxygen motion strongly dominates in modes above 262 cm$^{-1}$[36]. The Fe atoms responsible for ferromagnetism are mainly involved in modes of in-between range. The shift in vibrational frequencies higher than 212 cm$^{-1}$ corresponds to the displacement of Fe and O ions in FeO$_6$ octahedra as a result of disorder induced by La and Nd substitution which can affect the Fe–O–Fe bond angle. The substitution of La and Nd on Bi site affect the Bi-O bond distance significantly which is reflected as a large shift in the vibrational frequency of 170 cm$^{-1}$ mode. In the BLNFO film, La and Nd substitution at Bi site leads to the change in structure from rhombohedral to monoclinic because structural change always accompanies the change of Bi–O covalent bonds, including the bond length and the bond angle.

The Raman spectra with spectroscopic line broadening are closely related to the increase in various defects including mismatch between thin film and substrate, lattice vacancies, and local lattice disorders in the films [34, 37, 38]. The observed variation of the Raman line width along the growth direction of BFO films indicates that the lattice vacancies density is the dominant mechanism in influencing the microstructure of BFO films rather than the lattice mismatch between substrate/buffer and film [39].

### 3.3. Dielectric properties

Figure 10 exhibits the frequency dependence of relative dielectric constant ($\varepsilon_r$) and dielectric loss (tan δ) of BFO and BLNFO films. The $\varepsilon_r$ slightly increases with decreasing the frequency form 100 Hz to 1 MHz for both the films. The measured value of $\varepsilon_r$ are 140 and 280, and tan δ are 0.05 and 0.02 at 1KHz frequency for BFO and BLNFO films, respectively. This indicates that the La and Nd co-doping in BFO films leads to the increase in dielectric constant ($\varepsilon_r$) and reduces the dielectric loss. Figure 10(b) reveals the frequency dependent conductivity behaviour of BFO and BLNFO films. The conductivity values at 100 Hz are 1.9 ×10$^{-13}$ and 1.2×10$^{-13}$, and at 1MHz are 8.5×10$^{-11}$ and 4.5×10$^{-11}$ S/cm for BFO and BLNFO films respectively. It is well known that in ferroelectric films, under the applied electric field, space charge tend to accumulate at the interface between metal electrode and film which leads to the increase in relative dielectric constant ($\varepsilon_r$) and increase in dielectric loss (tan δ) at low frequencies. In case of BFO, the conduction mechanism is related with hopping of electrons from $Fe^{2+}$ to $Fe^{3+}$ when oxygen vacancies are present in the lattice. However, the co-substitution of La and Nd on A-site in BFO can suppress the formation of oxygen vacancies. The smaller number of oxygen vacancies lead to the lower density of space charge and hence, lower conductivity in ferroelectric films. It indicates that in BLNFO films, co-substitution of La and Nd can effectively improve the dielectric properties of BFO films.

**3.4. Magnetic properties**

The magnetic hysteresis loops were measured by employing SQUID with the magnetic field parallel to the film plane. Figure 11 shows magnetization versus field curves measured at 300 K for BFO and BLNFO films. Both the samples demonstrate ferromagnetic behaviour at room temperature with the saturation magnetization ($M_s$) of 8 and 20 emu/cm$^3$ for BFO and BLNFO films respectively. It indicated enhanced magnetization in films as compared to the bulk BFO and BLNFO compounds [18] which is due to the destruction of cycloid spin structure in thin films. The increased $M_s$ from 8 to 20emu/cm$^3$ might be due to the suppression of inhomogeneous spin structure or change in canting angle induced by substitution of La and Nd in BFO. It has been reported by Huang *et al* [15] that the saturation magnetization of BFO increased by Nd doping upto 10% and further increase of Nd reduces the $M_s$ value. The $M_s$ value decreases for 15 % Nd because of the decreased $Fe^{2+}$ ions together with enhanced second phase and increased defects. With the increase of Nd composition, the competition between the increasing structural distortion and the monotonically decreasing $Fe^{2+}$ fraction is the main cause for the $M_s$ going through a maximum at *x*=0.1. It has also been reported by Lee *et al* [10] that the saturation magnetization of these BFO-based epitaxial films increased with the degree of La modification. In the present work co-doped BFO films exhibit structural phase transition from R3c to nearly monoclinic phase (as analysed by Raman spectroscopy) that can leads to the destruction of the inhomogeneous spin structure. This structural distortion in BLNFO film releases the locked magnetization and enhances the macroscopic magnetization.

The BLNFO film shows a step like switching of the magnetization (as marked in fig. 11) in M-H loop that is attributed to the presence of antiferromagnetic phase along with induced ferromagnetic phase by substitution of La and Nd. Similar behaviour was also observed by S. Saxin *et al* which shows the sudden jump in the magnetic moment which is due to the presence of mixed phase structure [40].

## 4. Conclusion

In conclusion, BFO and BLNFO were successfully grown by pulsed laser deposition. XRD studies shows that the out of plane lattice parameter of the film has been decreased by doping with La and Nd. Raman analysis provide strong evidence that the BFO/SRO/STO film have rhombohedral R3c structure and doped BFO film undergoes structural transition from rhombohedral to monoclinic. Magnetic properties found to be improved upon doping might be due to the suppressed spatially non uniform spin structure and / or increased canting angle. Doped film possesses enhanced magnetic properties resulted from the structural distortion.

**Acknowledgement**

The authors would like to acknowledge Mr. Pankaj Pandey (UGC-DAE CSR, Indore) for help in magnetic measurements and Shushmita Bhartiya for help in AFM measurements.

**Caption of Figures**

Fig. 1: X-ray diffraction results of $BiFeO_3$ and $Ba_{0.8}La_{0.1}Nd_{0.1}FeO_3$ film grown over $SrRuO_3$ buffered $SrTiO_3$ (001) substrate. The inset shows enlarged portion of overlapped diffraction pattern for $BiFeO_3$ and $Ba_{0.8}La_{0.1}Nd_{0.1}FeO_3$ films.

Fig. 2: (a) Phi scan diffraction pattern of BFO based films and STO substrate on (022) plane (b) Reciprocal space mapping of the $BiFeO_3/SrRuO_3/SrTiO_3$ along (022) reflection.

Fig. 3: AFM micrograph of pure $BiFeO_3$ and doped $Ba_{0.8}La_{0.1}Nd_{0.1}FeO_3$ films deposited over $SrRuO_3$ buffered $SrTiO_3$ substrate.

Fig. 4: FESEM micrograph of $BiFeO_3$ and $Ba_{0.8}La_{0.1}Nd_{0.1}FeO_3$ films deposited over $SrRuO_3$ buffered $SrTiO_3$ substrate. The inset of fig. 3 shows dimensional image over 100 micron area.

Fig.5: Comparison of unpolarized Raman spectra of $SrTiO_3$ single crystal substrate, $BiFeO_3/SrRuO_3/SrTiO_3$ at room temperature. The inset of figure shows fitted $BiFeO_3$ peak (solid line) at 72.11 $cm^{-1}$ and $SrTiO_3$ peak (stars) at 78.68 $cm^{-1}$.

Fig. 6: Room temperature Raman spectra of $BiFeO_3$ and $Ba_{0.8}La_{0.1}Nd_{0.1}FeO_3$ $SrRuO_3/SrTiO_3$ films. Lines are drawn as a guide for eyes.

Fig.7: Low temperature Raman spectra of BFO (a) and BLNFO (b) films recorded in unpolarized geometry.

Fig.8: Representative multiple peak fitted spectra of BFO (a) and BLNFO (b) films.

Fig.9: Polarization dependent room temperature Raman spectra of (a) $BiFeO_3/SrRuO_3/SrTiO_3$ and $Ba_{0.8}La_{0.1}Nd_{0.1}FeO_3$ $SrRuO_3/SrTiO_3$ films; (b) STO substrate.

Fig.10: (a) Frequency dependence of dielectric constant ($\epsilon_r$) and dielectric loss (tan $\delta$) for the BFO and BLNFO films; (b) Frequency dependence of conductivity for BFO and BLNFO films.

Fig.11: Room temperature magnetization vs hysteresis curves for $BiFeO_3$ and $Ba_{0.8}La_{0.1}Nd_{0.1}FeO_3$ films.

**Table 1:** Comparison of Raman active modes and symmetry assignment of BiFeO$_3$: the present work with earlier reported results.

| Ref.[7] BFO/STO(001) $R3C$ | | Ref.[41] BiFeO$_3$ single crystal | | Ref [39] BFO/STO(001) $R3C$ | | Present BFO film | |
|---|---|---|---|---|---|---|---|
| Symmetry assignment | Frequency assignment | Symmetry assignment | Frequency assignment | Symmetry assignment | Frequency assignment | Symmetry assignment | Frequency assignment |
| A | 77 | E | 75 | E | 76 | E | 72 |
| A | 142 | E | 81 | A | 140 | A | 139 |
| A | 176 | E | 132 | A | 172 | A | 173.8 |
| A | 221 | A | 145 | A | 217 | A | 219 |
| E | 279 | A | 175.5 | E | 262 | E | 277 |
| E | 359 | A | 222.7 | E | 275 | E | 355 |
| E | 369 | E | 263 | E | 307 | E | 378 |
| E | 473 | E | 276 | E | 345 | A | 473 |
| E | 530 | E | 295 | E | 369 | E | 526 |
| E | 615 | E | 348 | E | 429 | E | 569 |
| | | E | 370 | A | 470 | E | 616 |
| | | E | 441 | E | 521 | | |
| | | A | 471 | E | 613 | | |
| | | E | 523 | | | | |
| | | A | 550 | | | | |

**Table 2.**

| Scattering geometry | R(R3c) (C$_{3v}$) | T(P4mm) (C$_{4v}$) | M(B$_b$) (C$_s$) |
|---|---|---|---|
| N (Raman) | 4A$_1$+9E | 3A$_1$+B$_1$+4E | 13 A$'$+14A$''$ |
| Z(XX)Z̄ | A$_1$ , E | A$_1$ , B$_1$ | A$'$ |
| Z(XY)Z̄ | E | No modes | A$''$ |

**FIGURE-1**

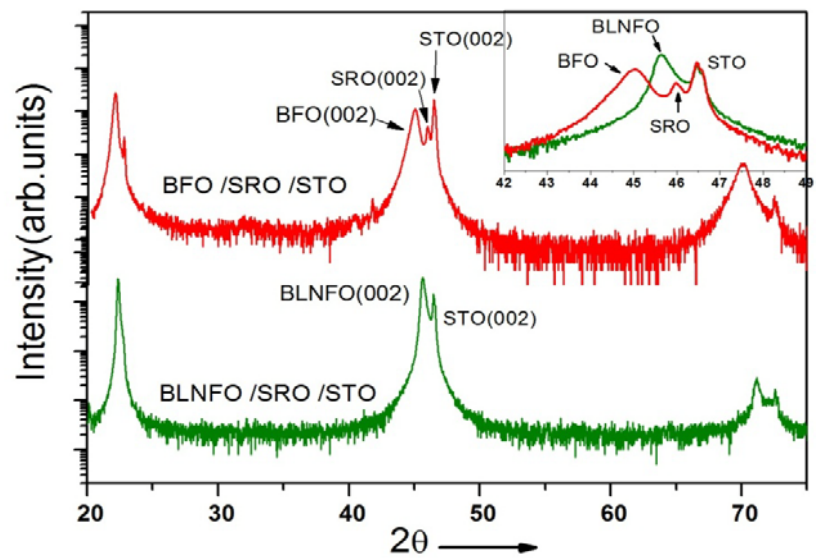

**FIGURE-2**

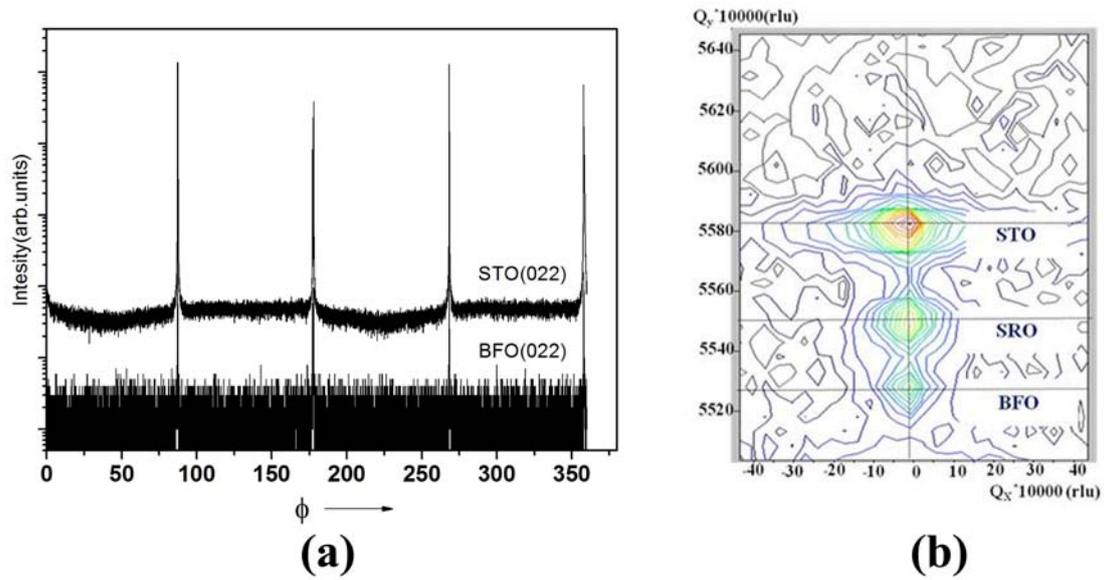

**FIGURE-3**

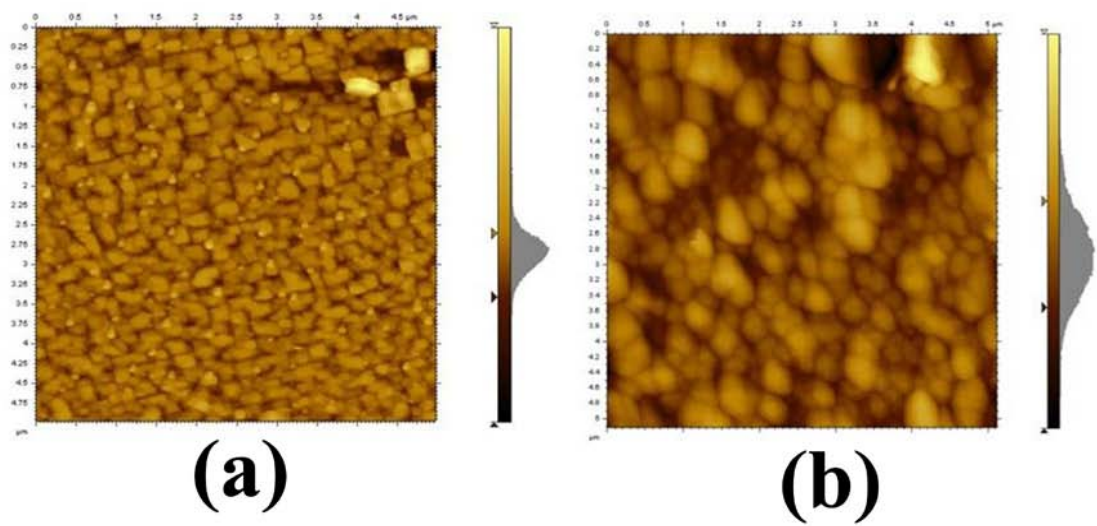

**FIGURE-4**

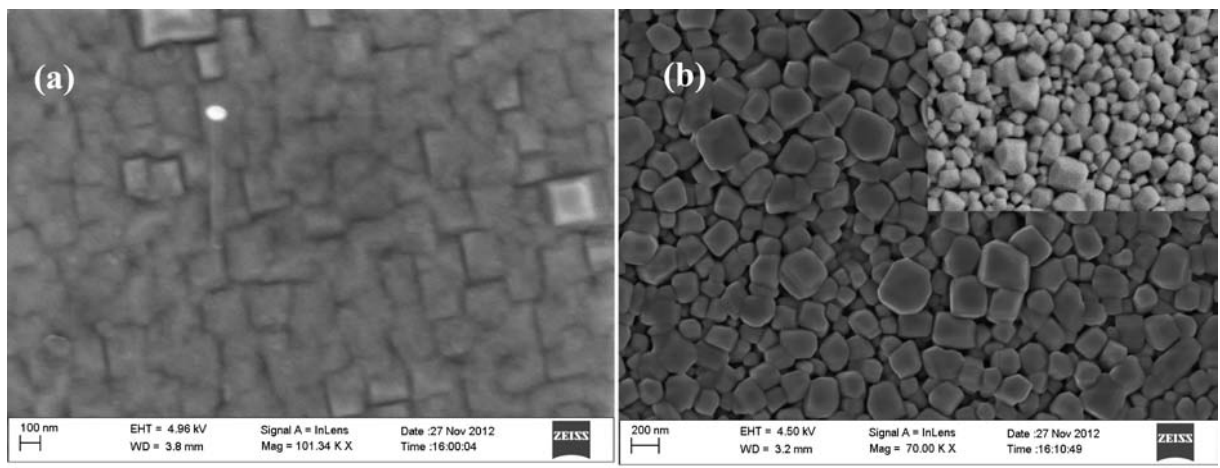

**FIGURE-5**

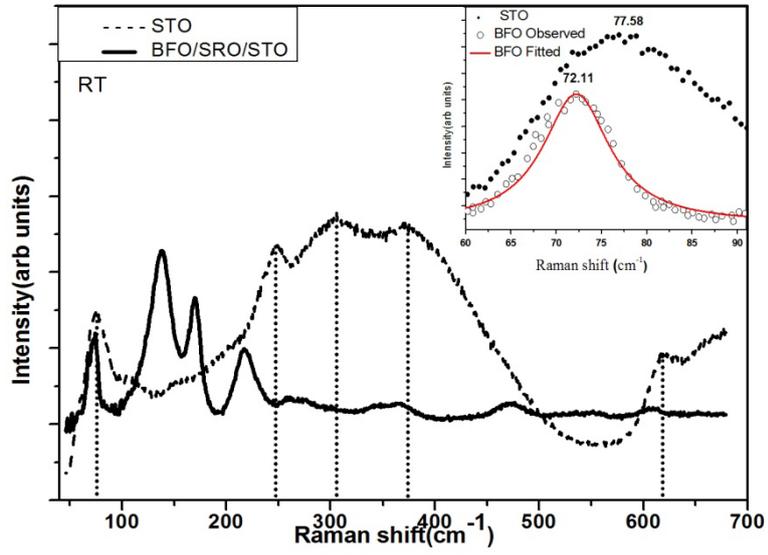

**FIGURE-6**

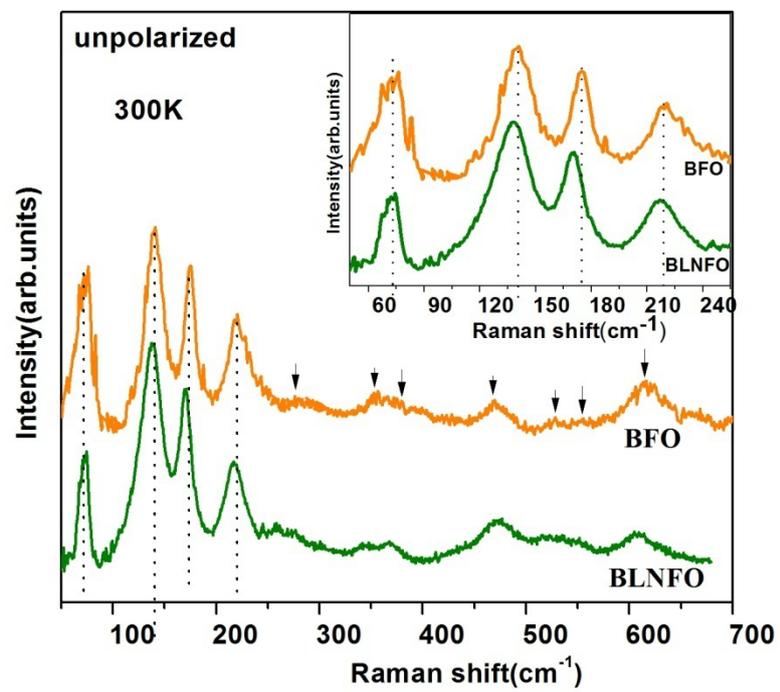

**FIGURE-7**

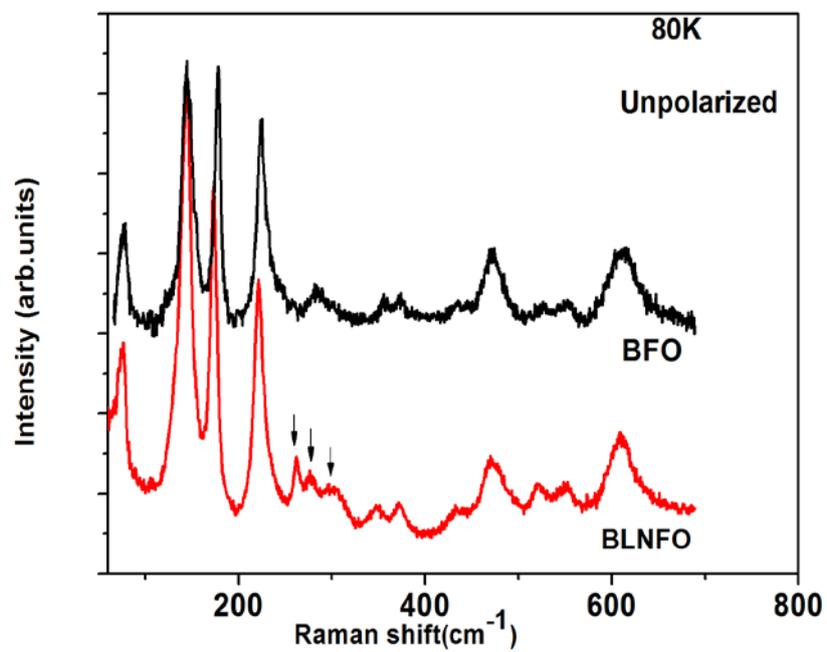

**FIGURE-8**

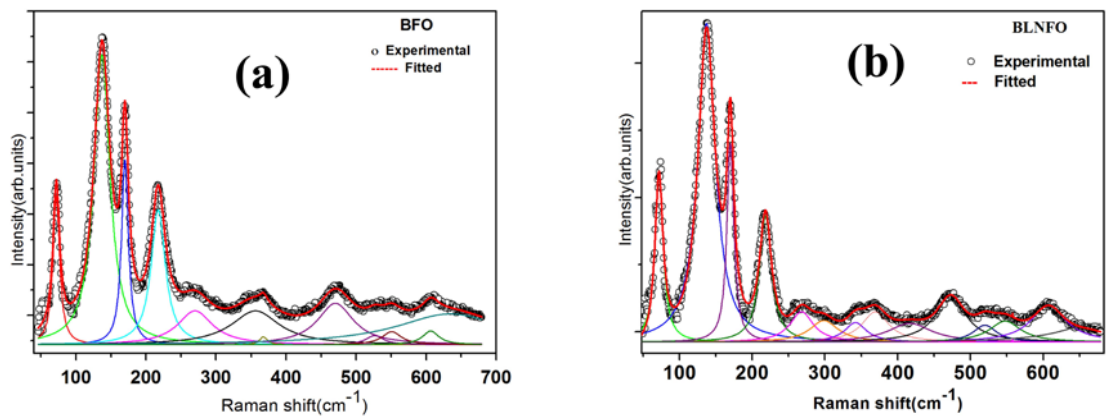

*FIGURE-9*

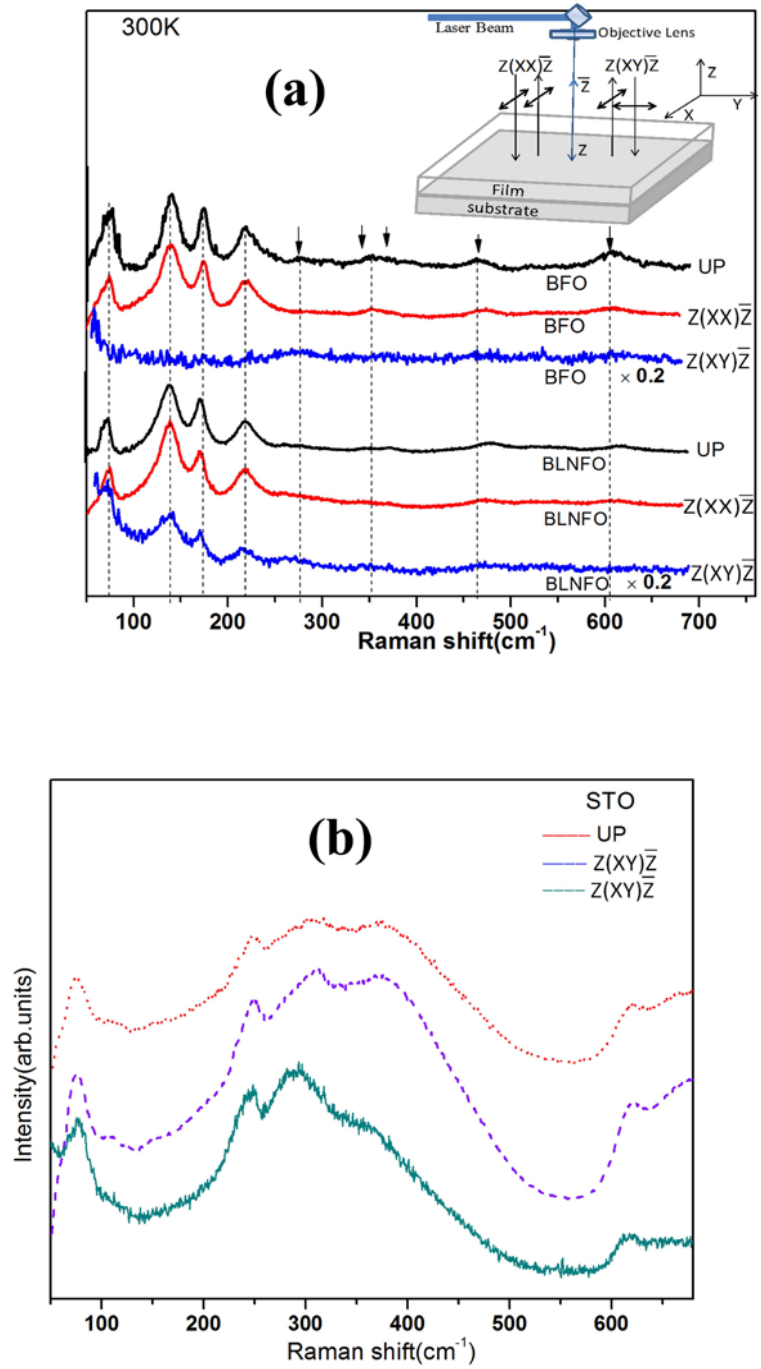

*FIGURE-10*

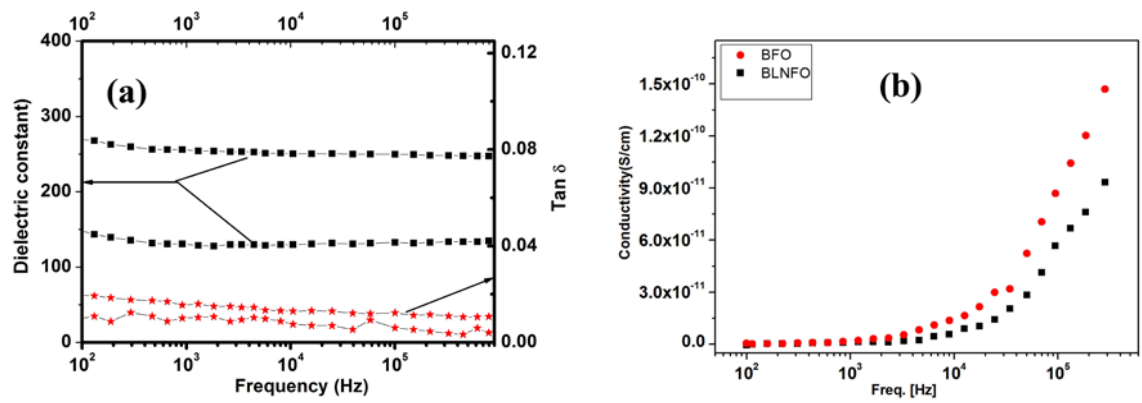

*FIGURE-11*

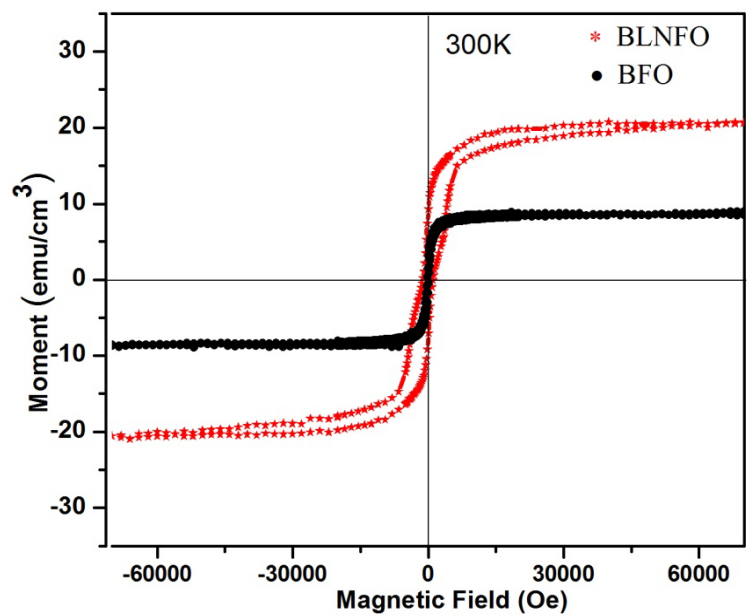